\documentclass[useAMS,usenatbib]{mn2e}

\newcommand{\phref}[3]{\textcolor{#1}{\href{#2}{#3}}}

\usepackage[dvips]{graphicx}

\newcommand{\mygraphics}[2]{\includegraphics[#1]{Images/#2}}

\usepackage{txfonts}
\usepackage{color}
\usepackage{setspace}
\usepackage{multirow}
\usepackage{hyperref}

\definecolor{corxcol}{rgb}{0.0,0.4,1.0}
\newcommand{\corx}[1]{#1}

\newcommand{\flab}[1]{\label{fig:#1}}

\newcommand{\slab}[1]{\label{sec:#1}}

\newcommand{\fref}[1]{Figure \ref{fig:#1}}

\newcommand{\cref}[1]{chapter \ref{chap:#1}}
\newcommand{\sref}[1]{section \ref{sec:#1}}

\newcommand{\elston}{{\em Elston-84} }
\newcommand{\dubner}{{\em Dubner-96} }
\newcommand{\nvss}{{\em NVSS-93} }

\definecolor{paulcol}{rgb}{0.4,0.1,0.6}

\title[Kinematics of the interaction between the SS433 jets and the
W50 nebula]{Probing the history of SS\,433's jet kinematics via Decade-resolution radio observations of W\,50}
\author[P.\,T. Goodall,  K.\,M. Blundell and S.\,J. Bell Burnell]{Paul T. Goodall$^{1}$\thanks{E-mail:
ptg@astro.ox.ac.uk (PTG)}, Katherine M.\ Blundell$^{1}$\thanks{E-mail: kmb@astro.ox.ac.uk (KMB)} and S.\ Jocelyn Bell Burnell$^{1}$\thanks{E-mail: jocelyn@astro.ox.ac.uk (SJBB)}\\
$^{1}$Department of Physics, University of Oxford, United Kingdom.}
\begin{document}

\date{Accepted 1988 December 15. Received 1988 December 14; in original form 1988 October 11}


\maketitle

\label{firstpage}

\begin{abstract}
We present the results of a kinematical study of the W\,50 nebula using high resolution radio observations from the Very Large Array (VLA) spanning a 12-year period, sampled in 1984, 1993 and 1996.  We conduct a careful analysis of the proper motions of the radio filaments across the W\,50 nebula at each epoch, and detect no significant motion for them during this period.  The apparent lack of movement in the radio filaments mandates either (i) a high degree of deceleration of SS\,433's jet ejecta in the W\,50 nebula, or (ii) that the lobes of W\,50 formed a long time ago in SS\,433's history, during a jet outburst with appreciably different characteristics to the well-known precessing jet state observed in SS\,433 at the present day.  We discuss the possible scenarios which could explain this result, with relevance to the nature of SS\,433's current jet activity.      
\end{abstract}


\section{Introduction}
The spectacular interaction between the relativistic jets of the
microquasar SS433 and the supernova remnant (SNR) W\,50, has been captured at
several epochs by radio telescopes that are sensitive to both its
angular extent (approximately two degrees across the sky) as well as to its
fine structure on arcsecond scales.  The connection between microquasar and SNR has been confirmed, as the distance to this structure is now very well known via two distinct methods.  In the first instance, a distance to SS\,433 of $5.5 \pm 0.2$\,kpc was determined by examination of light-travel time induced aberrations of the
arcsec-scale jets of SS\,433 \citep{kmb04}.  The second method is based upon the
detection of neutral hydrogen, in both absorption and in emission, at a high Galactic 
rotation speed corresponding to the same distance of 5.5\,kpc to W\,50 \citep{lockman07}.  Accurate distances to objects within our Galaxy are still rare, but
when such objects are identified their angular
sizes translate directly into physical sizes, proper motions convert nicely to velocities, and this information enables a quantitative assessment of the system energetics.

The central engine in SS\,433 is known to eject jet material with a mean speed of $\sim$0.26c as determined from optical spectroscopy \citep{eikenberry2001} and this is found to be consistent with proper motions of the jet knots using long-baseline radio interferometry \citep{Vermeulen,Paragi} with a standard deviation in the jet speed of $0.01c$ \citep{kmb04}.  \corx{The recent review of \citet{fabrika2004} provides a detailed description of SS\,433's behaviour and a comprehensive summary of different types of observation of this system.}

\corx{Although this paper focuses upon radio observations, evidence for the SS\,433-W\,50 interaction has been reported in several other wavelength regimes.  Optical maps of W\,50 \citep{zealey1980,boumis2007} show bright filaments approximately 30$^{\prime}$ to the east and west of SS\,433 that are suggested to be excited by the relativistic beams of SS\,433.  Broadband X-ray observations \citep{safiharb1997,brinkmann2007} show emission peaking just radially inwards of the extremities of the radio protrusions \citep[e.g.\ ][fig.\ 3]{blundellhirst2011} with hard non-thermal X-ray emission at 35$^{\prime}$ from SS\,433, and a general softening of the X-ray spectrum with increasing distance from SS\,433.  The X-ray emission originates near to the mean precession axis, well within the 20$^{\circ}$ semi-angle cone traced out by the precessing jets.   This has led to suggestions that SS\,433's jets are being refocused back toward the mean jet axis \citep{peter1993}, or that the X-ray emission arises due to reflective hydrodynamic shocks propagating along the symmetry axis because of the SNR-jet interaction \citep{velazquez2000}.  However, more recent hydrodynamic simulations of the SS\,433-W\,50 interaction show a new mechanism for hydrodynamic refocusing of conical jets \citep{ptg2010}, where W\,50's nebular morphology and SS\,433's refocused jets are both reproduced to a good approximation.   X-rays have been detected from SS\,433 itself by the Chandra satellite; on the basis of some of these observations \citet{migliari2002} suggest in-situ reheating of the baryonic components, while \citet{migliari2005} interpret the variability of the arcsec-scale X-ray jets as evidence of a second, faster outflow connected with the jets.  Variability and deviation from the standard kinematic model for the radio and optical jets in SS\,433 have been analysed and considered by \citet{kmb05} and \citet{kmb07}.}

Whilst the kinematics of SS\,433's jets have been generally well-studied near the point of jet-launch at both optical and radio wavelengths, the relativistic propagation of jet material at the outermost periphery has not yet been observed.  Indeed, any deceleration of the jet material, or the point at which deceleration happens, has yet to be established.       

The SS\,433-W\,50 system is amenable to numerical simulations, and the results of a wide range of hydrodynamic models are presented in a companion paper \citep{ptg2010}.  In the present paper, we present a detailed examination of three VLA radio images of W\,50 spanning 12 years.  We aim to ascertain whether meaningful constraints can be placed on the proper motion of the persistent radio features (the filamentary structure in W\,50) where the jets have impinged and penetrated through the supernova remnant (SNR) shell.  In particular, we aim to identify any evidence for the deceleration of the jets at the furthest point they are observed from the microquasar nucleus.

\section{Data Analysis}

\subsection{VLA Data Reduction}
For the datasets used in this analysis, the image published by
\citet{dubner98} was kindly made available to us by Gloria Dubner, and the NVSS image was downloaded from the NVSS ``postage stamp server'' \citep{condon1998}.
The data originally observed by \citet{elston84} were retrieved from the NRAO VLA data archive.  These data comprised nine telescope pointings,
observed at 1.4\,GHz in the VLA's most compact configuration, D-array,
and were reduced individually using standard procedures in AIPS.  The
individual pointings were then sequenced together using bespoke code
written in PerlDL.  For clarity we refer to these observations by source and observation year according to their respective FITS headers, as \elston ({\em date-obs}: 12/08/1984),  \nvss ({\em date-obs}: 15/11/1993), and \dubner ({\em date-obs}: 19/08/1996) respectively.

\subsection{Image Preparation}
\label{sec:imageprep}

Three epochs of W\,50 observations from the VLA in L-band and D-array configuration, provide us with maps of very similar resolution (the VLA synthesised beamwidth\footnote{http://www.vla.nrao.edu/astro/guides/vlas/current/node10.html} in this configuration is 44\arcsec) and size, with differences being due to the restoring beam used for each observation and the fact that each image is comprised of a series of mosaicked pointings.  An interactive kinematics pipeline was written by PTG using the Perl Data Language\footnote{The Perl Data Language (PDL) has been developed by K. Glazebrook, J. Brinchmann, J. Cerney, C. DeForest, D. Hunt, T. Jenness, T. Luka, R. Schwebel, and C. Soeller and can be obtained from http://pdl.perl.org.} (PDL, \cite{glazebrook97}) to extract the accurate fitted coordinates of the SNR filaments from the radio observations.  The pipeline performs a series of image preparation steps to ensure accurate image matching before kinematics can be measured, and the basic stages are outlined below.

The images were scaled-up from their original sizes (by a factor of at least 5) such that the image sizes and pixel scales approximately matched, using a high-order polynomial interpolation method to ensure accurate resampling.  Each image was then normalised such that the brightest pixel (always centred upon the unresolved source SS\,433) has a value of unity.  An accurate astrometric correspondence between the three observations was created using a select group of 15 bright point sources common to all three images (see Figures B.1 to B.3 in Appendix B).  This was done by fitting Gaussians to the point source centroids, and each fit was inspected by eye to ensure positional reliability.  The averaged coordinates $C$=($\bar{x}_{\rm c}$\,,\,$\bar{y}_{\rm c}$) of all 15 point sources (the centre of mass if all point sources are assigned an equal mass) was chosen as the principal reference point for each image:
\begin{equation}
{\rm CRPIX1} = \bar{x}_{\rm c} = \frac{1}{n} \sum_{i=1}^{n} x_{\rm i}~,
\end{equation}
and
\begin{equation}
{\rm CRPIX2} = \bar{y}_{\rm c} = \frac{1}{n} \sum_{i=1}^{n} y_{\rm i}~,
\end{equation}
where $n$ is the number of point sources used, and the error associated with any one measured centroid position is therefore minimised.  The 6-character capitalised keywords (CRPIX1 etc) refer to the standard FITS header keywords defined by \citet{Wells1981}.  We define the east centre $E$=($\bar{x}_{\rm E}$\,,\,$\bar{y}_{\rm E}$) as the average coordinates of all the point sources to the east of SS\,433 (point sources 2-9), and west centre $W$=($\bar{x}_{\rm W}$\,,\,$\bar{y}_{\rm W}$) as the average coordinates of the remaining point sources including SS\,433 (point sources 10-15 and 1) such that:
\begin{equation}
   \bar{x}_{\rm e} = \frac{1}{a} \sum_{i=2}^{9} x_{\rm i}~,
\end{equation}

\begin{equation}
   \bar{y}_{\rm e} = \frac{1}{a} \sum_{i=2}^{9} y_{\rm i}~,
\end{equation}

\begin{equation}
   \bar{x}_{\rm w} = \frac{x_{1}}{b} + \frac{1}{b}\sum_{i=10}^{15} x_{\rm i}~,
\end{equation}

\begin{equation}
   \bar{y}_{\rm w} = \frac{y_{1}}{b} +  \frac{1}{b}\sum_{i=10}^{15} y_{\rm i}~,
\end{equation}
where $a=8$ and $b=7$ are the number of sources used for the east and west sides respectively.  The separation between the east centre ($E$) and west centre ($W$) then defines an accurate image pixel scale for each image:

\begin{equation}
  \delta x =  \bar{x}_{\rm w} -  \bar{x}_{\rm e}~,
\end{equation}

\begin{equation}
  \delta y =  \bar{y}_{\rm w} -  \bar{y}_{\rm e}~,
\end{equation}

\begin{equation}
{\rm CDELT1} = {\rm CDELT2} =  \frac{1}{\sqrt{\delta x^2 +  \delta y^2}}.
\end{equation}

 The angle $\theta$ made by the vector $\overrightarrow{EW}$ (joining the east and west centres) with the horizontal axis is indicative of any residual rotation that must be performed to match the images:

\begin{equation}
   \theta = \arctan \frac{\delta y}{\delta x}.
\end{equation}

\begin{figure}
\centering
\mygraphics{width=\columnwidth}{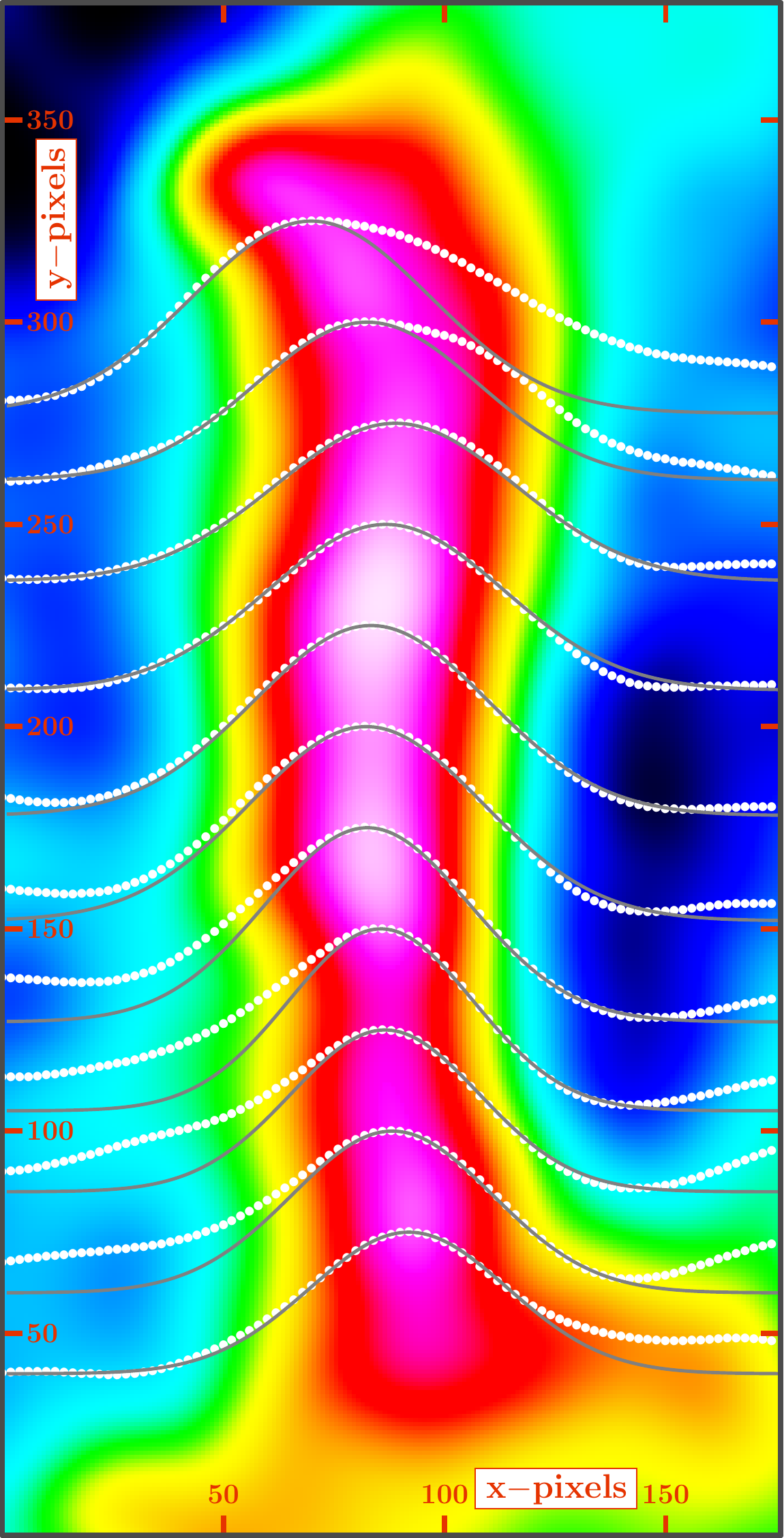}
\caption{\protect\centering \normalsize An example of the Gaussian centroid fitting to determine the filament mean positions along its length.  The filament has been rotated such that its longest axis is along the vertical, and the Gaussian curves are then fit horizontally line by line, before rotating the coordinates back to the image frame.  Note that only every 25th sampling is shown, and only every second datapoint per intensity profile.  Difficulties arise when the structure of the filament is more extended, causing the horizontal intensity profile to diverge from a Gaussian shape.  To compensate for this, each Gaussian curve is fit individually to determine a reliable range for the fit, and each fit is inspected by-eye to ensure a good match to the peak-position (the wings of the Gaussian are not important here).  The white points represent the data, and the grey lines indicate the best fit to each line of data shown.  Each Gaussian curve here has been overlaid upon the image, so that its peak aligns with the y$-$position from which it was taken.}
\flab{gaussfit}
\end{figure} 

The \elston and \nvss images were scaled and rotated to match the new header values of the \dubner image, and the Gaussian fitting to the 15 point source centroids was repeated to establish the correct header information for the newly transformed images, followed by a convolution to match both of \elston and \nvss to the slightly coarser resolution of $\theta_{\rm res}=$56\arcsec of the \dubner image.  Finally, the images were matched to within 0.01 pixels by performing an image product maximisation based upon the 15 point sources.  A final transformation was applied using a high-order polynomial interpolation, and the image FITS headers were updated one last time by fitting Gaussian centroids to the point sources in the matched images\footnote{An animation sequencing the final images is available here:  \phref{blue}{http://www-astro.physics.ox.ac.uk/~ptg/RESEARCH/research.html}{\mbox{$http://www$-$astro.physics.ox.ac.uk/$$\sim$$ptg/RESEARCH/research.html$}}}.

Throughout the analysis, non-linear corrections in astrometry were not accounted for, because the linear approximation is satisfactory over the degree scales involved.

\subsection{Velocity resolution}
\slab{vres}

The bright filaments in the images generally have a relatively high signal-to-noise and the Gaussian curves we fit to the cross sections of the filaments are well sampled across a given reliable range (determined by inspection, detailed in \S \ref{sec:kin}). Under these conditions, the accuracy with which one can determine the centroid position from a Gaussian fit is typically much finer than the Gaussian FWHM, often by a factor of 10 or more depending on the signal-to-noise.  Hence, the Gaussian centroid-fitting accuracy $x_{\rm res}$ can be written as 
\begin{equation}
   x_{\rm res} = \mu_{\rm fit}\,\frac{\theta_{\rm res}}{3600}\frac{\pi}{180} \,d_{\rm SS433} = 0.149\, {\rm pc} = 0.487\, {\rm ly}
\end{equation}
where $d_{\rm SS433}=5500$\,pc is the distance from us to SS\,433 and W\,50, and $\mu_{\rm fit}=0.1$ represents the modest factor of 10 times better accuracy in determining the centroid position, as compared with the image resolution.  The time baseline $t_{\rm base}$ between the \elston and \dubner images is 12 years and 4 days (with three of those years being leap years) giving a total of 4387 days, or of $t_{\rm base}=12.01$ years based upon the 365.25-day year.  The velocity resolution in units of the speed of light is then:
\begin{equation}
   v_{\rm res} = \frac{x_{\rm res}}{t_{\rm base}} = 0.0405\,c.
\end{equation}

This velocity resolution is a factor of 6.5 times smaller than the measured speed of SS\,433's jets of $v_{\rm jet}=0.2647c$ \citep{eikenberry2001} close to their launch point, and so the kinematics of the movement of filaments associated with the jets (or turbulence caused by the jets) should be measurable, unless significant deceleration has occurred.

\begin{figure*}
\begin{minipage}{\textwidth}
\centering
\mygraphics{width= \textwidth}{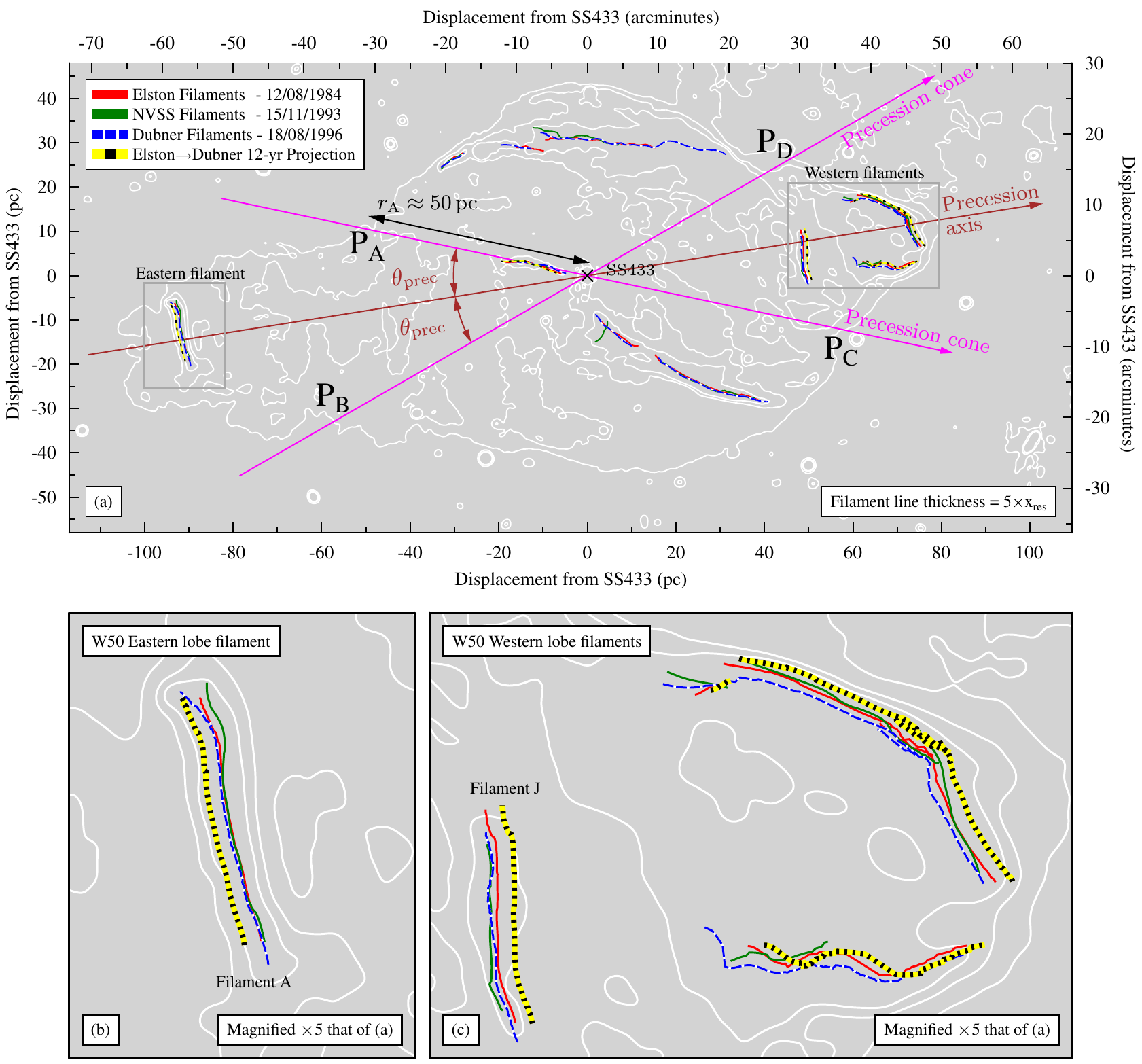}
\caption{\protect\centering \normalsize {\bf(a)}: The filament positions of each of the three observations are overlaid upon a contour map of the \dubner observation.  The yellow-black dashed lines represent the pseudo-dataset generated by projecting the \elston filament positions 12 years into the future, assuming a speed of 0.26c in the direction radially away from  SS\,433.  The brown arrow indicates SS\,433's jet precession axis, which is also W\,50's symmetry axis.  The magenta arrows indicate the edges of SS\,433's precession cone in the plane of the sky, based upon the cone semi-angle at the current epoch of $\theta_{\rm jet} = 20.92^{\circ}$ \citep{eikenberry2001}.  The points P$_{\rm A}$, P$_{\rm B}$, P$_{\rm C}$, and P$_{\rm D}$ indicate the locations where SS\,433's current jet configuration would penetrate the W\,50 nebula.  Note the remarkable alignment between the bright filament nearest to SS\,433 and the north-east jet cone line at image coordinates $\sim$(-10,+3) parsecs.  The radial distance $d_{\rm A}$ from SS\,433 to point P$_{\rm A}$ is approximately 50 parsecs.  The errors in the fitted positions of the filaments are of the order of (or smaller than) the thickness of the filament lines plotted on the figure. {\bf(b \& c)}:  A zoom-in on each of the eastern and western filaments.  Note that filaments A and J which are the most prominent from the radio images, are almost perpendicular to the precession axis.}
\flab{filaments}
\end{minipage}
\end{figure*} 

\subsection{Measuring Filament Kinematics in W\,50}
\label{sec:kin}
The ten brightest filament-like structures common to all three images were selected and labelled A-H as shown in Figures \ref{fig:elston84}, \ref{fig:nvss93}, and \ref{fig:dubner96}.  \corx{Although other common filamentary features are present between the three images, only the ten brightest were used based upon the criteria of having a a signal-to-noise ratio greater than five, and being at least four times as long as wide, in all three images.  Also, the three images are not of the same quality; the \dubner image is by far the deepest, followed by the \elston image.  NVSS was a survey designed to map the bright radio sky quickly, and so the \nvss image of W\,50 is much lower fidelity due to the lower integration time on target.  For this reason, the filament brightnesses vary considerably across the three images, and the \nvss image was the limiting factor in choosing suitably bright filaments.  Filament B was later rejected from the \nvss dataset\footnote{The lack of Filament B from the NVSS observation does not affect any of the conclusions of this paper.} during the analysis by the pipeline, as not being bright enough determine its position accurately.}  Using the interactive kinematics pipeline (as described in \ref{sec:imageprep}), the longest axis of each filament was determined and rotated to the vertical, such that the Gaussian curve fitting could be applied to the horizontal direction, to each pixel-row of the filament region.  The fitting process could not be automated\footnote{This involved over 2000 Gaussian fit curves per image.} due to extended structural variations in the background emission around each filament, and each Gaussian curve fit was inspected by-eye to ensure the best-fit was measured for the peak position of the filament and not to background structure which contributed to the Gaussian wings.  An example of the Gaussian fit curves to Filament-A is given in \fref{gaussfit}, and the complete set of filaments from each image is given in appendix B.

\section{Results}

The filament positions for each of the three epochs are overlaid together upon a contour image of the \dubner observation in Figure \ref{fig:filaments}.  A pseudo-dataset was generated using the data from the \elston observation, by projecting those filament positions 12 years into the future (based upon SS\,433's current jet speed at the point of launch) in order to compare the expected pseudo-filament positions with the \dubner data.  This projection was only applied to filaments that fall within SS\,433's precessional jet cone, as material far from the mean jet axis will not have interacted directly with the jets and cannot be expected to propagate at similar speeds.  Inspection of Figure \ref{fig:filaments} reveals that the filament positions from each of the three observations are remarkably coincident, with little indication of any conclusive movement, whereas the projected filaments are noticeably displaced from the rest.  This raises two possibilities: 
\begin{enumerate}
\item either considerable deceleration of the jet ejecta has occurred between jet-launch and the filaments of the jet lobes, or 
\item the filaments in W\,50 have not interacted with the current jet (of speed $\sim$0.26c) observed to be active near the central engine SS\,433.
\end{enumerate}
A velocity map (Fig. \ref{fig:vmap}) taken from high resolution hydrodynamic simulations of the SS\,433-W\,50 interaction (detailed in \citet{ptg2010}) shows no signs of the required deceleration along the jet axis, and therefore supports the second possibility.

\section{Discussion}
The analysis reveals no discernible changes in the filament positions during the 12-year baseline covered by these observations. \corx{We therefore infer an upper limit to the proper motion of the filaments equal to the velocity resolution of this investigation: $v_{\rm res}=0.0405c$ or $\mu=0.466^{\prime\prime}\,yr^{\rm -1}$}.  

One conclusion to draw from this (as from (i) above) is that considerable deceleration of the jet ejecta has occurred through interactions with the interstellar medium (ISM) present along the path of the jet lobes of W50, causing a reduction in the speed of the material in the lobes of W50.  This deceleration must be sufficient to reduce the speed of the jet material to below the detectable velocity resolution ($v_{\rm res}=0.0405c$) of this kinematic study.  These data would imply a lower limit of $\sim 85\%$ kinetic energy loss in the jets, and this energy would have been transferred to the ISM of the lobes. However, high resolution hydrodynamic simulations of the jets in SS\,433 \citep{ptg2010} reveal that, for any reasonable combination of jet mass-loss rate in the range $10^{-5}-10^{-4}\,{\rm M}_{\odot}\, {\rm yr}^{-1}$ and ISM density in the range $0.1-1$ particles cm$^{-3}$, a {\em persistently active} jet of speed $v_{\rm jet}=0.26c$ does not suffer such extreme deceleration (see \fref{vmap}).  We stress however, that this model is based upon a {\em persistent} jet, whereby kinetic energy is continually injected from the central engine in SS\,433 to replenish the kinetic energy (and momentum) lost to the jet, and transferred to the surroundings via interactions with the ISM.

In order to create a possible scenario that is in agreement both with the kinematics data presented here and the hydrodynamical simulations of the companion paper \citet{ptg2010}, we must invoke the idea of intermittent, rather than persistent, jet behaviour in SS\,433.  The case for discontinuous jet activity in SS\,433's history can be realised through assimilation of the relevant facts:
\begin{enumerate}
\item \corx{Our hydrodynamic simulations \citep{ptg2010} follow the evolution of a supernova explosion, until the SNR reaches 45\,pc in radius, equal to that of the circular region of the W\,50 nebula.  The expansion timescale (or equivalently, the SNR age) depends upon the background ISM density, where low ISM densities facilitate a more rapid SNR expansion and vice-versa.  For the ISM density in the range 0.1 to 1 particle(s) cm$^{-3}$, we find a corresponding SNR age of approximately 20 to 40 kyrs respectively.}

\item   \corx{The magenta lines in Fig. \ref{fig:filaments} indicate the edge of SS\,433's jet precession cone in the plane of the sky.  By extrapolating the trajectory followed by the jet when in the plane of the sky, one would expect the jets to penetrate the W\,50 nebula at the points P$_{\rm A}$, P$_{\rm B}$, P$_{\rm C}$, and P$_{\rm D}$ (henceforth ``the jet penetration points''), and alter the morphology of the nebula significantly.  However, the radio observation of \citet{dubner98} (see top panel of \fref{dubner96}) shows a surprisingly gradual morphological transition from the central circular region to the east and west lobes of the nebula, with no obvious extrusions at the jet penetration points.  The lack of extrusions at the jet penetration points can be explained in one of to ways, either:\vspace{0.2cm} \\{\bf [Scenario 1]} - SS\,433's jets have been somehow refocused back towards the precession axis, thus preventing any morphological deformation at the jet penetration points.  \vspace{0.2cm}\\{\bf [Scenario 2]} - SS\,433's current jet precession state has not been active long enough for jets travelling at the current speed of 0.26c to reach the jet penetration points.\vspace{0.2cm}}  

\item \corx{The jet refocusing hypothesis has been investigated by various authors (see for example \citet{eichler1983,Kochanek1990,peter1993}), and is a particularly appealing scenario because it could simultaneously explain the confinement of the X-ray emission \citep{brinkmann2007} to the region close to the jet precession axis.  However, our hydrodynamical simulations \citep{ptg2010} show that jet refocusing does indeed occur {\em but} via a different mechanism to those previously suggested.  We find that refocusing of the jet trajectory occurs via momentum exchange with the turbulent interstellar gas contained within the nebula, which has been agitated by the jet itself.  Although this refocusing mechanism does indeed provide an explanation for the confinement of X-ray emission near the jet precession axis (see Figure 8f and h of the companion paper \citep{ptg2010}), the side-effect of imparting momentum to the interstellar gas is to inflate the nebula, causing significant deformation at the jet penetration points (see Figure 8(e) and (g) of the companion paper \citep{ptg2010}).  Thus, the refocusing of SS\,433's hollow conical jets is unlikely to prevent nebular deformation at the jet penetration points marked on \fref{filaments}, and we can safely rule out scenario 1.}

\item \corx{According to \fref{filaments}, the nearest jet penetration point to SS\,433 is P$_{\rm A}$ at a distance of $r_{\rm A} \approx 50$pc.  Assuming a jet speed of 0.26c, the {\em ``in vacuo''} jet travel time from SS\,433 to P$_{\rm A}$ is $\sim$600 years.  The jets do not travel in a vacuum however, and our hydrodynamic simulations \citep{ptg2010} test the effects of various environmental properties (e.g. ISM density) upon the dynamics of jets with the same characteristics as SS\,433's jets (precession angle, jet speed etc).  For comparison, simulating SS\,433's jets propagating through an ISM density of 0.2 particles cm$^{-3}$, begins to have a noticeable impact on the SNR morphology at the jet penetration points approximately 3500 years after the jets are switched on.  Dynamical arguments can be used to estimate the likely age/duration of the current jet precession state at $\lesssim$3500 years\footnote{Note that at higher ISM densities than 0.2 particles cm$^{-3}$ the jets would require more than 3500 years to deform the nebula at the penetration points, and less than 3500 years at lower ISM densities, with a lower limit at the {\em``in vacuo''} timescale.}.  This is considerably less time than the SNR age given in (i), and therefore favours the possibility of several jet episodes previous to the currently observed one.}  

\item \corx{Our hydrodynamic simulations \citep{ptg2010} also tested the effects of varying the jet precession cone angle.  The results indicate that any non-zero precession cone angles produce wider jet lobes than could explain the W\,50.  Cylindrical jets are a special case of conical jets with a precession angle of zero, and only in this special case are lobes produced of the correct width when compared with the lobes of W\,50.  Since SS\,433's current jet activity features precession with a cone angle of approximately 20$^{\circ}$, it cannot be responsible for the lobes of W\,50, and this is further evidence in favour of a previous jet episode having precession cone angle close to zero.}

\item \corx{Filaments A and J from \fref{filaments} lie directly along SS\,433's jet precession axis.  The lack of movement from these filaments indicates an upper limit to the filament speed equal to the velocity resolution of this investigation, of $v_{\rm res}=0.0405c$.  This speed is much lower than SS\,433's current jet speed of 0.26c, and the velocity difference of 0.22c places a lower limit on the kinetic energy loss of jets, such that 85\% of the kinetic energy per unit mass is transferred from the jets to the surrounding ISM, between jet launch and reaching the extent of filaments A and J.  \fref{vmap} shows the velocity map from one of our hydrodynamic simulations of a persistently active jet with speed 0.26c.  The velocity map shows that the persistent jets maintain very close to their launch speed even once the reach the full extent of W\,50's lobes.  The simulation featured an ISM density of 0.2 particles cm$^{-3}$ which is a reasonable estimate for the ISM at the location of W\,50 within the Galaxy.  In fact, an untenably high ISM density would be required in the simulation, in order to explain the reduction in jet speed to below $v_{\rm res}=0.0405c$ in the jet lobes.  It is therefore unlikely that a persistent jet with speed 0.26c is responsible for the lobes of W\,50.  We propose two alternatives which would be consistent with the kinematics data: \vspace{0.2cm}\\(a) A previous jet episode with a lower jet speed (i.e $\lesssim v_{\rm res}$) could have created the jet lobes of W\,50.  This does not require an unfeasible level of deceleration in the jets. \vspace{0.2cm}\\(b) A previous jet episode of much shorter duration than the time required to produce the lobes, could be responsible.  Upon cessation of the past jet activity, the transfer of kinetic energy from the jets to the surrounding ISM would then (and only then, without further input of energy from the power source SS\,433) significantly decelerate the jet lobes, in agreement with the data presented here and from our hydrodynamic simulations \citep{ptg2010}.\vspace{0.2cm}\\Both of (a) and (b) require an additional jet episode dissimilar and previous to SS\,433's current jet activity.}
 
\item Intermittent jet activity is thought to be a common property of both microquasars and quasars \corx{\citep{nipoti2005}}, and jet flares are routinely observed in other microquasars such as Cygnus X-3 \citep{mioduszewski2001,mj2004}, also GRS1915-105 \citep{mirabel94} and V4641 \citep{hjell2000}.  For these microquasars however, the intermittency timescales are sub-year up to a few years, whereas we find hints of sub-millennium intermittency timescales for the SS\,433 system.
\end{enumerate}
\begin{figure*}
\begin{minipage}{\textwidth}
\centering
\mygraphics{width= \textwidth}{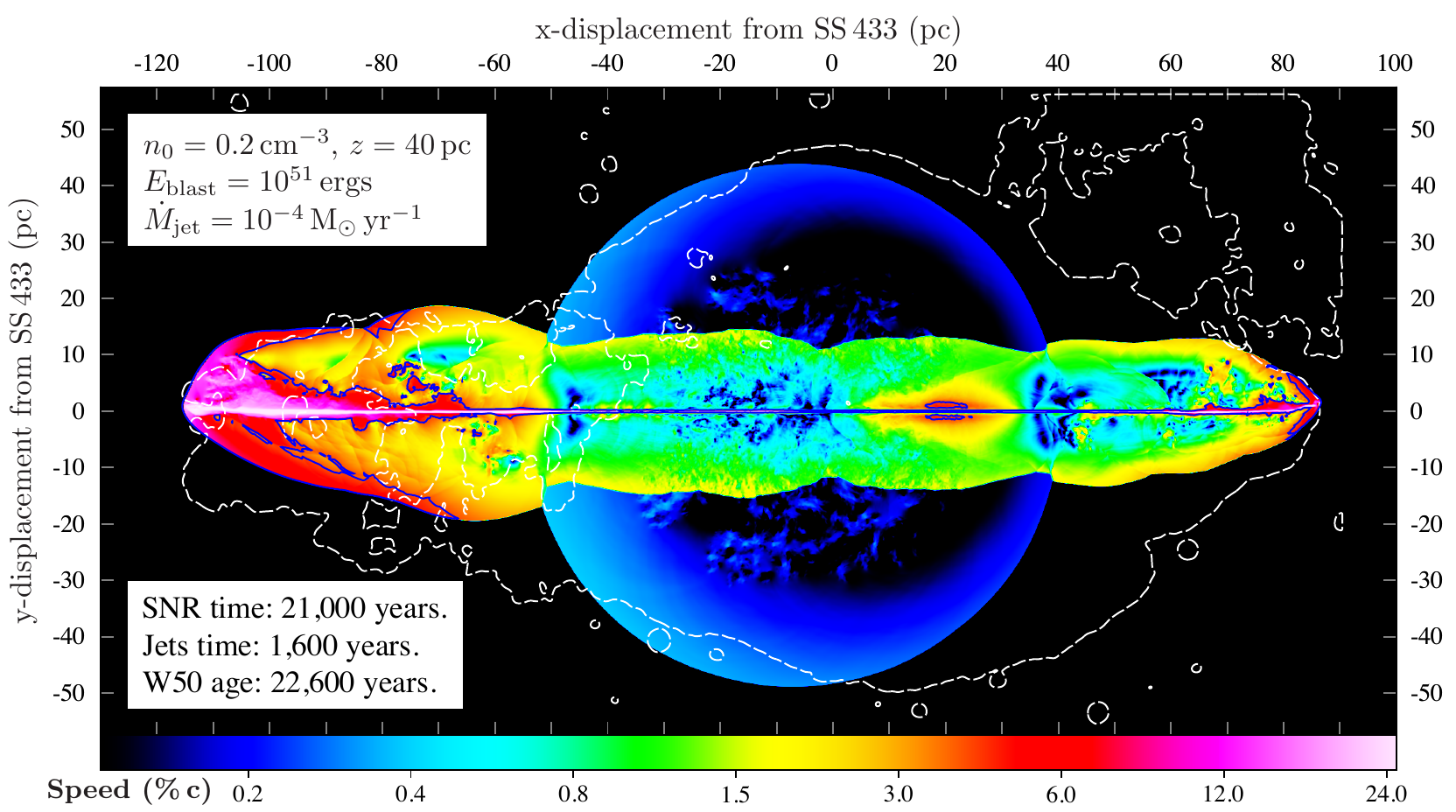}
\caption{\protect\centering We present a velocity map from high-resolution hydrodynamic simulations of the SNR-Jet interaction in SS\,433 (see \citet{ptg2010} for a detailed account of these simulations).  A contour map of the \dubner observation (white dashed line) has been overlaid upon the image to trace out the outline of the W\,50 nebula.  The above model can be summarised as a persistent cylindrical (zero precession cone angle) jet with speed 0.26c.  The jet is incident upon an evolved SNR (blue circular region) of radius $\gtrsim$40pc and age 21,000 years.  The background density profile ($n_{\rm 0}=0.2 $ particles cm$^{-3}$, $z=40$pc) has been optimised to match the observations of the east-west asymmetry in W\,50, and it takes 1,600 years for the jets to reach the extent of W\,50's lobes.  The colour bar indicated the speed of the moving gas as a percentage of the speed of light, and the blue contour (solid line) indicates jet material which is travelling faster than the velocity resolution calculated in \sref{vres}.  The harpoon-like light-pink/white regions (particularly evident in the eastern lobe) shows material which is travelling very close to the speed of the jet (0.26c) at the launch-point, which is a consequence of continuous ejection of kinetic energy from the persistent jet.}
\flab{vmap}
\end{minipage}
\end{figure*}

\section{Conclusions}
In contrast to SS\,433's reputation as a persistent jet source, we propose that the wealth of observational data on SS\,433 and W\,50 is consistent with SS\,433 having experienced (at least) two distinctly different states of jet activity.  First, the well-studied current jet state features $t_{\rm jet}=163$ day precession with cone angle $\theta_{\rm jet} = 20.92^{\circ}$ and jet speed $v_{\rm jet}=0.2647c$ \citep{eikenberry2001} and is observable both through the \corx{Doppler} shifting of spectral emission lines and high-resolution radio interferometry.  Second, a further state of jet activity which occurred some time in the past with the same axis of symmetry as the current jet system and either a zero (or very small) jet precession cone angle, or a precession period significantly longer than the duration of the jet activity, observable indirectly through the formation of the lobes of W\,50.  We note that the observations of Cyg X-3's jet-flaring event in 1997 were reported to have $v_{\rm 1997}\geq0.81c$ and precession cone angle $\psi_{\rm 1997}\lesssim 12^{\circ}$ \citep{mioduszewski2001}, and just four years later a speed $v_{\rm 2001}=0.63c$ and precession cone angle $\psi_{\rm 2001}=2.4^{\circ}$ were observed by \citet{mj2004}. The kinematic results presented herein, underline the need for further investigation via hydrodynamic simulations, of the possible intermittent jet activity in SS\,433's past.

\section*{Acknowledgments}

P. T. G. would like to thank the Science and Technology Facilities
Council for a D.Phil Studentship.  K. M. B. thanks the Royal Society
for a University Research Fellowship.   The National Radio Astronomy Observatory is a facility of the National Science Foundation operated under cooperative agreement by Associated Universities, Inc.  We warmly thank Sebasti\'an P\'erez for suggesting improvements to this manuscript.  

Portions of the analysis presented here made use of the Perl Data Language (PDL) developed by K. Glazebrook, J. Brinchmann, J. Cerney, C. DeForest, D. Hunt, T. Jenness, T. Luka, R. Schwebel, and C. Soeller and can be obtained from http://pdl.perl.org. PDL provides a high-level numerical functionality for the Perl scripting language \citep{glazebrook97}.

\bibliographystyle{mn2e} 
\bibliography{my_mnras_refs} 

\begin{thebibliography}{}

\bibitem[\protect\citeauthoryear{{Blundell} \& {Bowler}}{{Blundell} \&
  {Bowler}}{2004}]{kmb04}
{Blundell} K.~M.,  {Bowler} M.~G.,  2004, ApJ, 616, L159

\bibitem[\protect\citeauthoryear{{Blundell} \& {Bowler}}{{Blundell} \&
  {Bowler}}{2005}]{kmb05}
{Blundell} K.~M.,  {Bowler} M.~G.,  2005, ApJ, 622, L129

\bibitem[\protect\citeauthoryear{{Blundell}, {Bowler} \&
  {Schmidtobreick}}{{Blundell} et~al.}{2007}]{kmb07}
{Blundell} K.~M.,  {Bowler} M.~G.,    {Schmidtobreick} L.,  2007, AAP, 474, 903

\bibitem[\protect\citeauthoryear{{Blundell} \& {Hirst}}{{Blundell} \&
  {Hirst}}{2011}]{blundellhirst2011}
{Blundell} K.~M.,  {Hirst} P.,  2011, ApJLett

\bibitem[\protect\citeauthoryear{{Boumis}, {Meaburn}, {Alikakos}, {Redman},
  {Akras}, {Mavromatakis}, {L{\'o}pez}, {Caulet} \& {Goudis}}{{Boumis}
  et~al.}{2007}]{boumis2007}
{Boumis} P.,  {Meaburn} J.,  {Alikakos} J.,  {Redman} M.~P.,  {Akras} S.,
  {Mavromatakis} F.,  {L{\'o}pez} J.~A.,  {Caulet} A.,    {Goudis} C.~D.,
  2007, \mnras, 381, 308

\bibitem[\protect\citeauthoryear{{Brinkmann}, {Pratt}, {Rohr}, {Kawai} \&
  {Burwitz}}{{Brinkmann} et~al.}{2007}]{brinkmann2007}
{Brinkmann} W.,  {Pratt} G.~W.,  {Rohr} S.,  {Kawai} N.,    {Burwitz} V.,
  2007, \aap, 463, 611

\bibitem[\protect\citeauthoryear{{Condon}, {Cotton}, {Greisen}, {Yin},
  {Perley}, {Taylor} \& {Broderick}}{{Condon} et~al.}{1998}]{condon1998}
{Condon} J.~J.,  {Cotton} W.~D.,  {Greisen} E.~W.,  {Yin} Q.~F.,  {Perley}
  R.~A.,  {Taylor} G.~B.,    {Broderick} J.~J.,  1998, \aj, 115, 1693

\bibitem[\protect\citeauthoryear{{Dubner}, {Holdaway}, {Goss} \&
  {Mirabel}}{{Dubner} et~al.}{1998}]{dubner98}
{Dubner} G.~M.,  {Holdaway} M.,  {Goss} W.~M.,    {Mirabel} I.~F.,  1998, AJ,
  116, 1842

\bibitem[\protect\citeauthoryear{{Eichler}}{{Eichler}}{1983}]{eichler1983}
{Eichler} D.,  1983, \apj, 272, 48

\bibitem[\protect\citeauthoryear{{Eikenberry}, {Cameron}, {Fierce}, {Kull},
  {Dror}, {Houck} \& {Margon}}{{Eikenberry} et~al.}{2001}]{eikenberry2001}
{Eikenberry} S.~S.,  {Cameron} P.~B.,  {Fierce} B.~W.,  {Kull} D.~M.,  {Dror}
  D.~H.,  {Houck} J.~R.,    {Margon} B.,  2001, ApJ, 561, 1027

\bibitem[\protect\citeauthoryear{{Elston} \& {Baum}}{{Elston} \&
  {Baum}}{1987}]{elston84}
{Elston} R.,  {Baum} S.,  1987, AJ, 94, 1633

\bibitem[\protect\citeauthoryear{{Fabrika}}{{Fabrika}}{2004}]{fabrika2004}
{Fabrika} S.,  2004, Astrophysics and Space Physics Reviews, 12, 1

\bibitem[\protect\citeauthoryear{{Glazebrook} \& {Economou}}{{Glazebrook} \&
  {Economou}}{1997}]{glazebrook97}
{Glazebrook} K.,  {Economou} F.,  1997, Dr. Dobbs Journal

\bibitem[\protect\citeauthoryear{{Goodall}, {Alouani-Bibi} \&
  {Blundell}}{{Goodall} et~al.}{2011}]{ptg2010}
{Goodall} P.~T.,  {Alouani-Bibi} F.,    {Blundell} K.~M.,  2011, MNRAS,
  accepted (arXiv:1101.3486)

\bibitem[\protect\citeauthoryear{{Hjellming}}{{Hjellming}}{2000}]{hjell2000}
{Hjellming} R.~M.,  2000, \apj, 544, 977

\bibitem[\protect\citeauthoryear{{Kochanek} \& {Hawley}}{{Kochanek} \&
  {Hawley}}{1990}]{Kochanek1990}
{Kochanek} C.~S.,  {Hawley} J.~F.,  1990, \apj, 350, 561

\bibitem[\protect\citeauthoryear{{Lockman}, {Blundell} \& {Goss}}{{Lockman}
  et~al.}{2007}]{lockman07}
{Lockman} F.~J.,  {Blundell} K.~M.,    {Goss} W.~M.,  2007, MNRAS, 381, 881

\bibitem[\protect\citeauthoryear{{Migliari}, {Fender} \&
  {M{\'e}ndez}}{{Migliari} et~al.}{2002}]{migliari2002}
{Migliari} S.,  {Fender} R.,    {M{\'e}ndez} M.,  2002, Science, 297, 1673

\bibitem[\protect\citeauthoryear{{Migliari}, {Fender}, {Blundell}, {M{\'e}ndez}
  \& {van der Klis}}{{Migliari} et~al.}{2005}]{migliari2005}
{Migliari} S.,  {Fender} R.~P.,  {Blundell} K.~M.,  {M{\'e}ndez} M.,    {van
  der Klis} M.,  2005, \mnras, 358, 860

\bibitem[\protect\citeauthoryear{{Miller-Jones}, {Blundell}, {Rupen},
  {Mioduszewski}, {Duffy} \& {Beasley}}{{Miller-Jones} et~al.}{2004}]{mj2004}
{Miller-Jones} J.~C.~A.,  {Blundell} K.~M.,  {Rupen} M.~P.,  {Mioduszewski}
  A.~J.,  {Duffy} P.,    {Beasley} A.~J.,  2004, \apj, 600, 368

\bibitem[\protect\citeauthoryear{{Mioduszewski}, {Rupen}, {Hjellming}, {Pooley}
  \& {Waltman}}{{Mioduszewski} et~al.}{2001}]{mioduszewski2001}
{Mioduszewski} A.~J.,  {Rupen} M.~P.,  {Hjellming} R.~M.,  {Pooley} G.~G.,
  {Waltman} E.~B.,  2001, \apj, 553, 766

\bibitem[\protect\citeauthoryear{{Mirabel} \& {Rodr{\'{\i}}guez}}{{Mirabel} \&
  {Rodr{\'{\i}}guez}}{1994}]{mirabel94}
{Mirabel} I.~F.,  {Rodr{\'{\i}}guez} L.~F.,  1994, \nat, 371, 46

\bibitem[\protect\citeauthoryear{{Nipoti}, {Blundell} \& {Binney}}{{Nipoti}
  et~al.}{2005}]{nipoti2005}
{Nipoti} C.,  {Blundell} K.~M.,    {Binney} J.,  2005, \mnras, 361, 633

\bibitem[\protect\citeauthoryear{{Paragi}, {Vermeulen}, {Fejes}, {Schilizzi},
  {Spencer} \& {Stirling}}{{Paragi} et~al.}{1999}]{Paragi}
{Paragi} Z.,  {Vermeulen} R.~C.,  {Fejes} I.,  {Schilizzi} R.~T.,  {Spencer}
  R.~E.,    {Stirling} A.~M.,  1999, New Astronomy Review, 43, 553

\bibitem[\protect\citeauthoryear{{Peter} \& {Eichler}}{{Peter} \&
  {Eichler}}{1993}]{peter1993}
{Peter} W.,  {Eichler} D.,  1993, \apj, 417, 170

\bibitem[\protect\citeauthoryear{{Safi-Harb} \& {Oegelman}}{{Safi-Harb} \&
  {Oegelman}}{1997}]{safiharb1997}
{Safi-Harb} S.,  {Oegelman} H.,  1997, \apj, 483, 868

\bibitem[\protect\citeauthoryear{{Vel{\'a}zquez} \& {Raga}}{{Vel{\'a}zquez} \&
  {Raga}}{2000}]{velazquez2000}
{Vel{\'a}zquez} P.~F.,  {Raga} A.~C.,  2000, \aap, 362, 780

\bibitem[\protect\citeauthoryear{{Vermeulen}, {Icke}, {Schilizzi}, {Fejes} \&
  {Spencer}}{{Vermeulen} et~al.}{1987}]{Vermeulen}
{Vermeulen} R.~C.,  {Icke} V.,  {Schilizzi} R.~T.,  {Fejes} I.,    {Spencer}
  R.~E.,  1987, \nat, 328, 309

\bibitem[\protect\citeauthoryear{{Wells}, {Greisen} \& {Harten}}{{Wells}
  et~al.}{1981}]{Wells1981}
{Wells} D.~C.,  {Greisen} E.~W.,    {Harten} R.~H.,  1981, A\&AS, 44, 363

\bibitem[\protect\citeauthoryear{{Zealey}, {Dopita} \& {Malin}}{{Zealey}
  et~al.}{1980}]{zealey1980}
{Zealey} W.~J.,  {Dopita} M.~A.,    {Malin} D.~F.,  1980, MNRAS, 192, 731

\end{thebibliography}

\newpage
\appendix 
\section{Additional details regarding the Image preparation}
\corx{
\subsection{Image Product Maximisation}
This method compares the positions of the 15 points sources in each image to accurately align the images.  This works because the product of two similar 2D Gaussians is greatest when their peaks are coincident.}

\corx{A mask is first applied to each image, to set the image pixels to zero everywhere except for a small circular region around each of the point sources.  Image product maximisation is then performed to align any two of the masked-images.  }

\corx{The second masked-image I$_{\rm M2}$ is shifted (resampled) by some small value $dx$ in the x-direction, and $dy$ in the y-direction.  The images are then multiplied together and the sum of the product is recorder, according to:}
\begin{equation}
S(dx,dy) = \sum_{x,y}  I_{\rm M1}(x,y)\, I_{\rm M2}(x+dx,y+dy)
\end{equation}
  
\corx{The function $S(dx,dy)$ is a 2D Gaussian (because the point sources being cross-correlated are also Gaussians).  The $(dx_{\rm p},dy_{\rm p})$ coordinate corresponding to the peak of $S(dx,dy)$ denotes the shift that must be applied to the second image in order for both images to be optimally aligned.}

\newpage 
\section{Fitting to the filamentary structure of the W\,50 nebula.}
The following images show the fits to the centroid positions of the filamentary structures common to each of the three VLA observations.

\begin{figure*}
\centering
\mygraphics{width=0.9\textwidth}{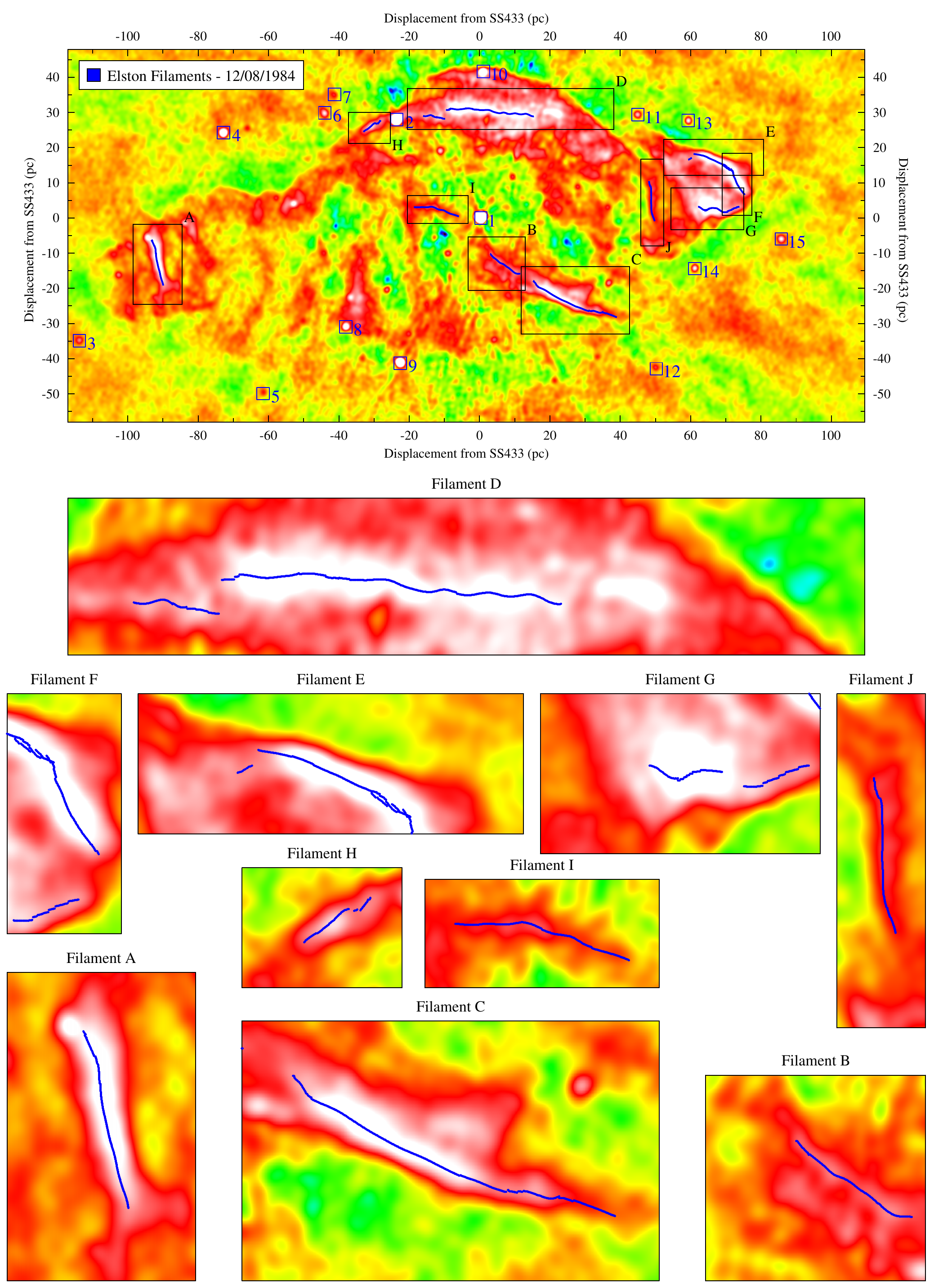}
\caption{\protect\centering The centroid positions are shown for the filaments of the \elston observation.}
\flab{elston84}
\end{figure*}

\begin{figure*}
\centering
\mygraphics{width=0.9\textwidth}{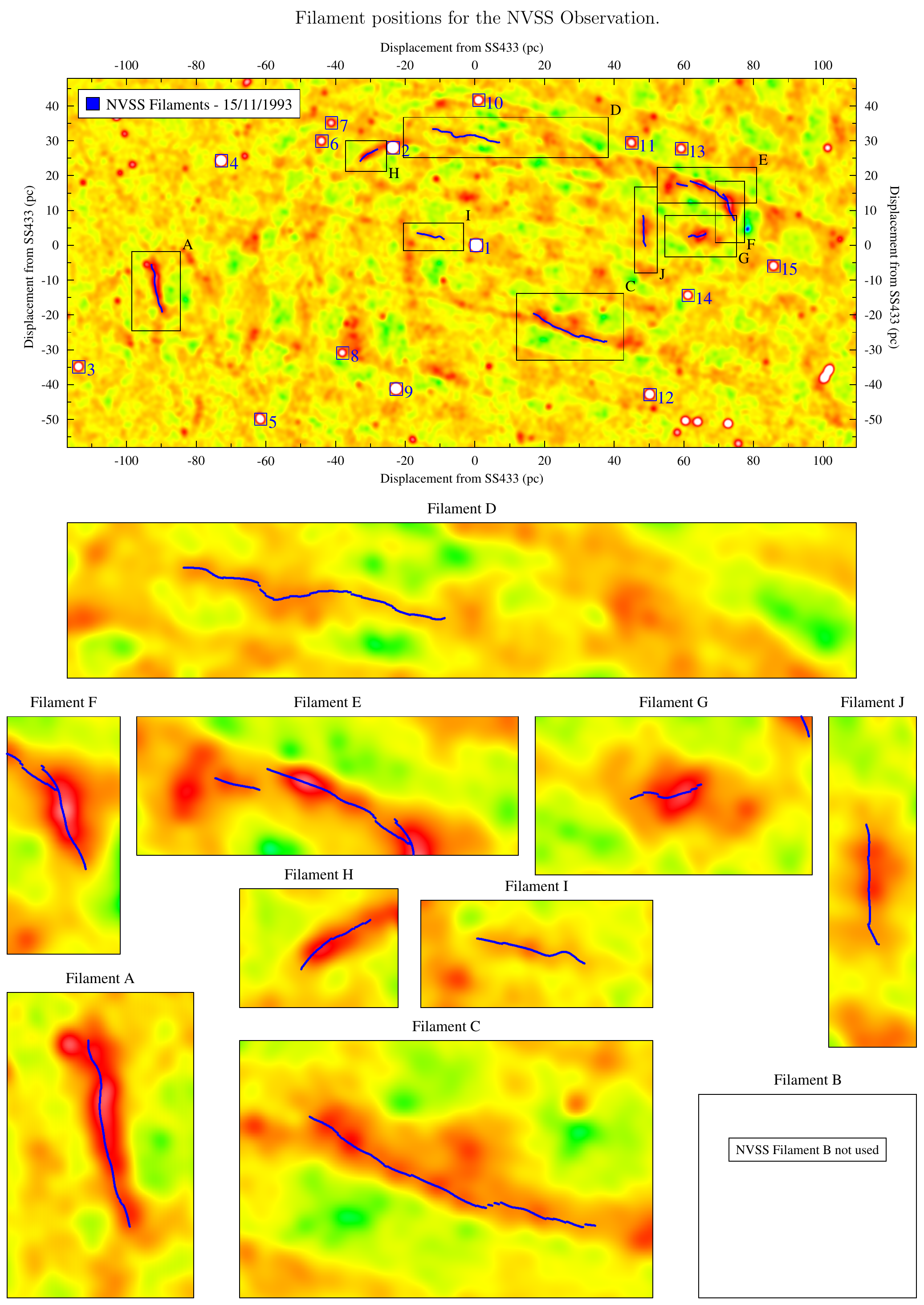}
\caption{\protect\centering The centroid positions are shown for the filaments of the \nvss observation.}
\flab{nvss93}
\end{figure*} 

\begin{figure*}
\centering
\mygraphics{width=0.9\textwidth}{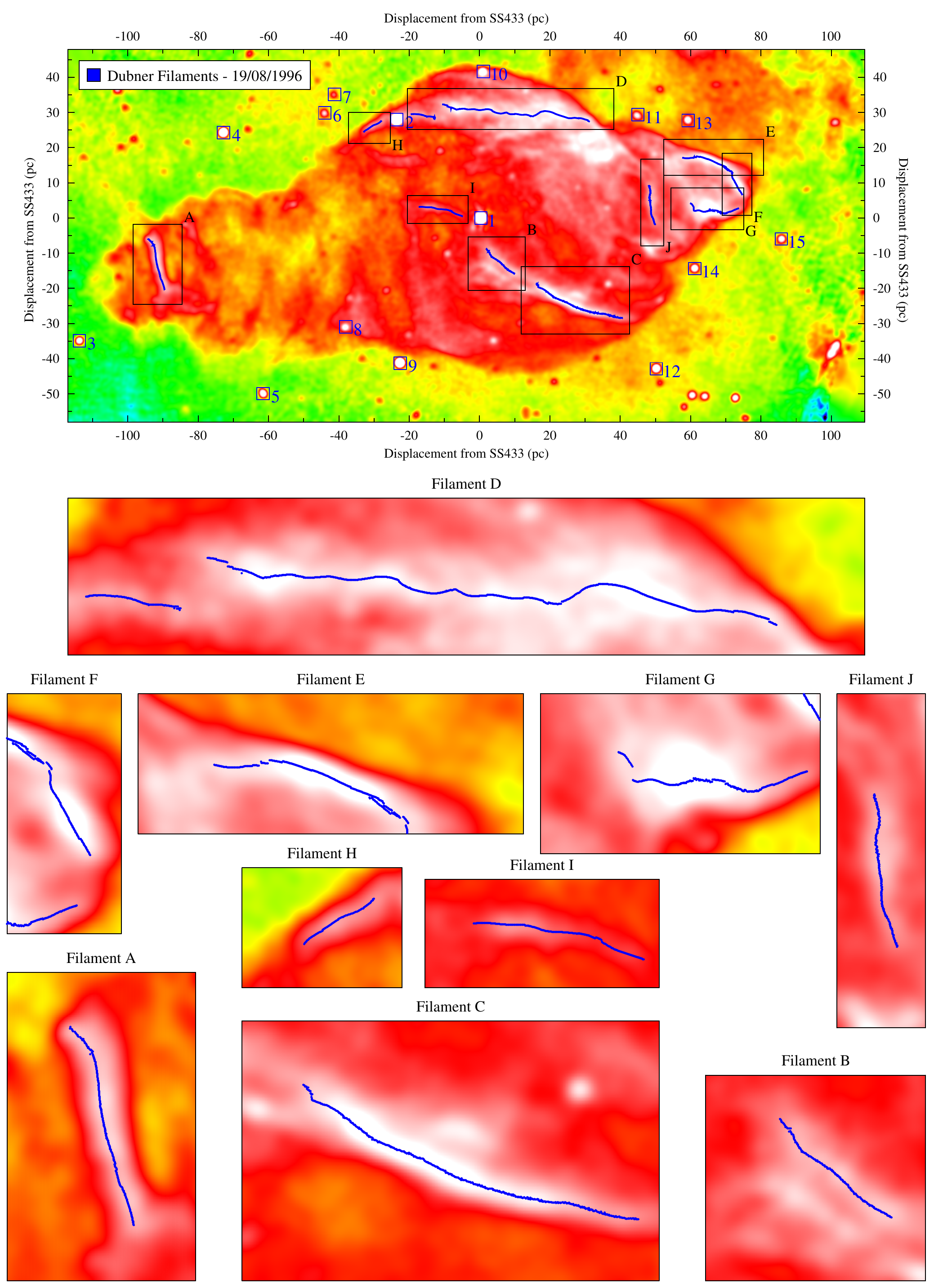}
\caption{\protect\centering The centroid positions are shown for the filaments of the \dubner observation.}
\flab{dubner96}
\end{figure*}

\end{document}